\documentclass[twoside,a4paper,11pt]{proceedings}
\usepackage{graphicx}
\usepackage{hyperref}
\usepackage{movie15}
\usepackage{natbib}
\begin{document}
\pagenumbering{arabic}
\pagestyle{myheadings}
\thispagestyle{empty}
\vspace*{-1cm}
%{\flushleft\includegraphics[width=\textwidth,bb=58 650 590 680]{stamp.pdf}}
%{\flushleft\includegraphics[width=\textwidth,viewport=58 650 590 680]{stamp.pdf}}
{\flushleft\includegraphics[width=3cm,viewport=0 -30 200 -20]{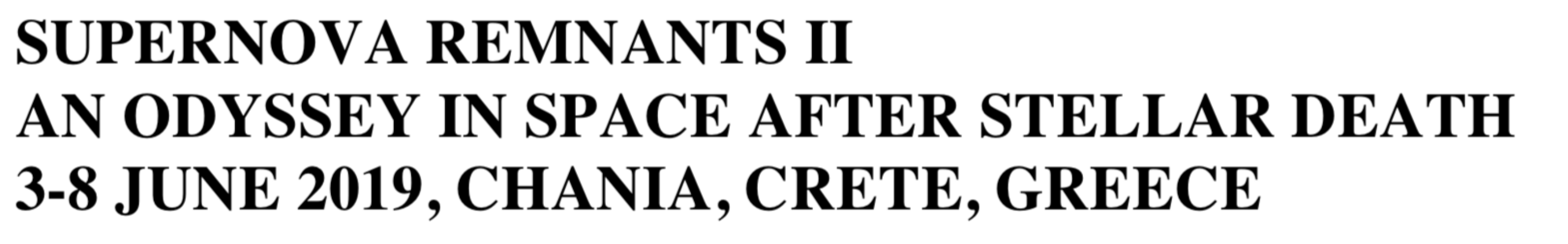}}
\vspace*{0.2cm}
\begin{flushleft}
{\bf {\LARGE
%%% TITLE of the paper. 
New X-ray observations toward PSR J1826$-$1256
}\\
\vspace*{1cm}
%%% Include here the LIST OF AUTHORS.
%%% Note that the last author has to be preceeded by an AND.
L. Duvidovich$^1$,
E. Giacani$^1$,
G. Castelletti$^1$
A. Petriella$^1$
and L. Supan$^1$
%,
%
% Do not delete next few lines
}\\
\vspace*{0.5cm}
%
%%% AFFILIATIONS LIST.
%%% and the AFFILIATIONS LIST. Note that one affiliation per line.
%%% Add as many affiliations as necessary. 
$^{1}$
CONICET-Universidad de Buenos Aires, Instituto de Astronom\'ia y F\'isica del Espacio (IAFE), Buenos Aires, Argentina. %\\
%$^{2}$
%Universidad de Buenos Aires, Facultad de Ciencias Exactas y Naturales, Buenos Aires, Argentina. \\
%$^{3}$
%Universidad de Buenos Aires, Ciclo Básico Común, Buenos Aires, Argentina.\\
%$^{4}$
%Universidad de Buenos Aries, Facultad de Arquitectura, Dise\~no y Urbanismo, Buenos Aires, Argentina.
%
% Do not delete next few lines
\end{flushleft}
% Headings
\markboth{
%%% Type the SHORT version of the paper t
X-ray toward PSR J1826$-$1256
}{
%%%  First Author \& Second Author   OR   First-author et al. 
%%%  First Author \& Second Author   OR   First-author et al. if the author list contains three or more authors.
Duvidovich et al.
}
\thispagestyle{empty}
\vspace*{0.4cm}
\begin{minipage}[l]{0.09\textwidth}
\ 
\end{minipage}
\begin{minipage}[r]{0.9\textwidth}
\vspace{1cm}
\section*{Abstract}{\small
%%% Type the ABSTRACT of your paper
We present the results of \textit{XMM-Newton} observations toward the pulsar PSR J1826$-$1256, which lies at 5$^{\prime}$.4 from the centroid of the TeV source HESS J1826$-$130. These data show an elongated nebula with the pulsar located in the northern border with the emission in the direction of the peak of the very high energy source. The spectral study of the X-ray emission shows a clear softening of the photon index with increasing distance from the pulsar. We discuss the connection among the pulsar, its nebula, and the HESS source.
%The XMM-Newton study of the region including PSR J1826−1256 reveals an elongated non-thermal X-ray
%emitting nebula with the pulsar located in the northern border and a tail towards the peak of the very high energy source. The spectrum is characterized by a power law. From our X-ray analysis we propose that HESS J1826−130 is likely produced by the PWN powered by PSR J1826−1256 via the inverse Compton mechanism.
\vspace{10mm}
\normalsize}
\end{minipage}
%%% BODY of the paper

\section{Introduction}

The pulsar PSR J1826$-$1256, one of the brightest radio-quiet $\gamma$-ray pulsars, is surrounded by diffuse and weak X-ray emission of nonthermal origin.
It was suggested that this emission detected with \textit{Chandra} could come from a pulsar wind nebula (PWN) powered by PSR J1826$-$1256, named G18.5$-$0.4 or ``Eel'' PWN \citep{Roberts07}.
PSR J1826$-$1256 is seen projected 5$^{\prime}$.4 away from the centroid of the TeV
source HESS J1826$-$130, which is a newly-unidentified extended TeV source,
previously hidden within the emission from the bright nearby PWN HESS
J1825$-$137 \citep{HESS18a}. In this paper, we present a new X-ray study with the aim to confirm the nature of the X-ray emission around PSR J1826$-$1256 and investigate its connection with HESS J1826$-$130 in order to unveil the origin of the $\gamma$-ray emission.

\section{Observations and Results}
\subsection{X-ray emission}

\begin{figure}[h]
\centering
\includegraphics[width=0.8\textwidth]{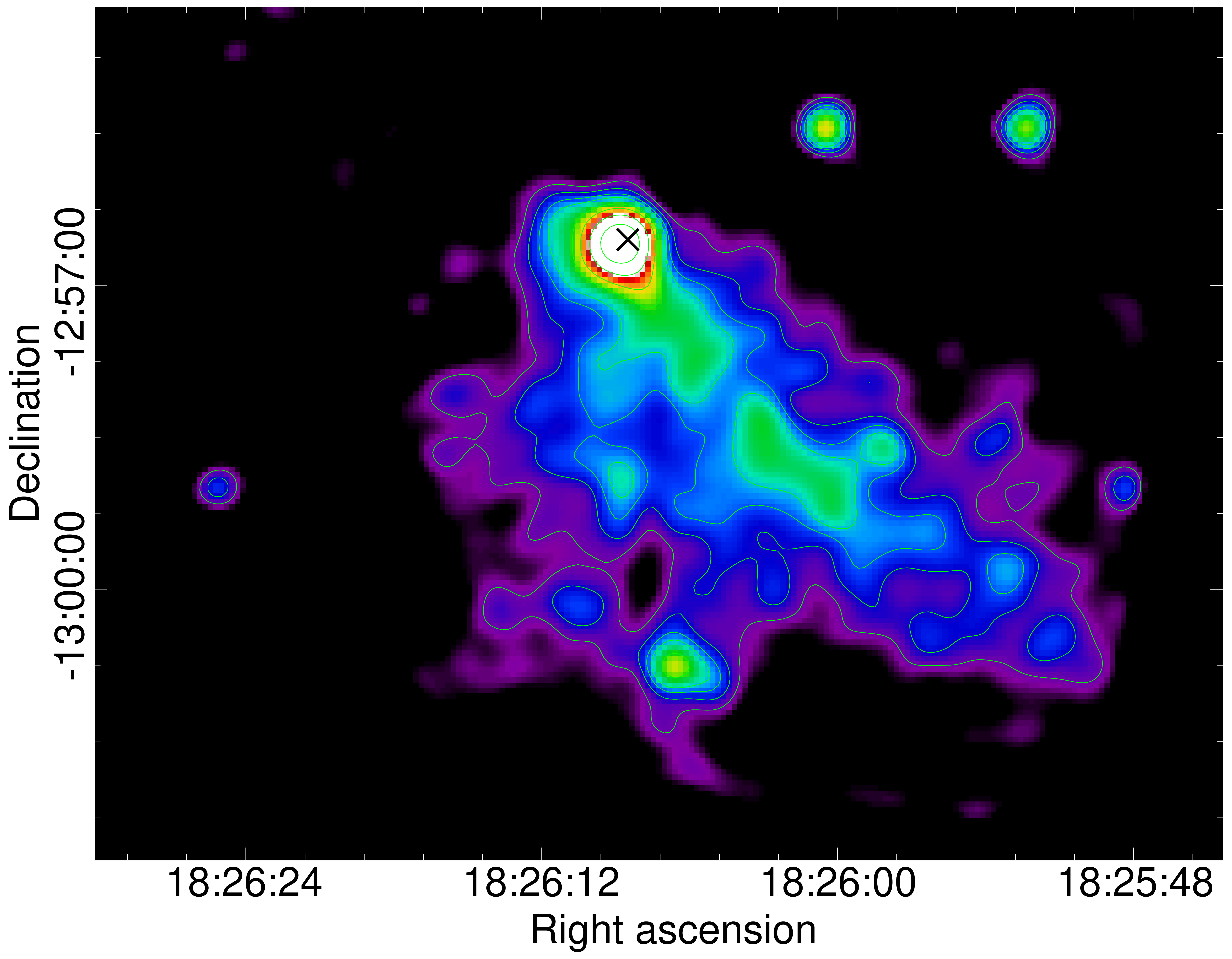}
\caption{First \textit{XMM-Newton} image in the direction to PSR J1826$-$1256 in the 1-7 keV energy band. The position of the pulsar is indicated with a cross sign, the contour levels are 5, 5.5, 6,  6.5, 7.5, 8.5, 10.5, and 14 cts/px.}
\label{espectro}
\end{figure}

The X-ray image and spectra toward PSR J1826$-$1256 were obtained by
processing unpublished archival \textit{XMM-Newton} data. The MOS1 and MOS2
cameras were set in the full-frame mode and hence mapped the full extension of
the nebula, while the PN camera was operated in the small-window mode, which
covers a small region around PSR J1826$-$1256.
The two MOS cameras were used for the analysis of the X-ray while the
three cameras for the spectral study of the pulsar. The data were processed
using software packages SAS 16.1.0 and Heasoft 6.22.1 following standard procedures.

%\subsection{Global analysis}

%In order to obtain the overall properties of the X-ray emission, the spectrum was extracted from the region delineated by the ellipse shown in the figure.

Figure \ref{espectro} shows the new \textit{XMM-Newton} image toward PSR J1826$-$1256 in the energy range 1 to 7 keV. This image reveals considerable new structures and diffuse emission not detected in the previous study performed on the basis of \textit{Chandra} data. The bulk of the emission comes from an elongated feature with an elliptical shape of about 6$^{\prime}$ $\times$ 2$^{\prime}$, which is brighter around the pulsar and extends toward the southwest in the direction of the centroid of the HESS 1826$-$130 TeV source. The elongated feature is surrounded by faint and diffuse emission that is more prominent toward the southeast.

To obtain the overall properties of the X-ray emission, the spectrum was extracted from the entire source.
The background was chosen from a circular region free of diffuse emission and excluding the point sources over it. The spectral points were simultaneously fit in the 1-7 keV energy band with an absorbed power-law model (\textit{wabs+powerlaw}). On the other hand, the spectrum of PSR J1826$-$1256 is also well fitted with an absorbed power-law model. 
We noted that the absorbing column density turns out to be similar to that obtained for the extended emission, as expected if they are associated sources.
%The fitted absorbing column density turns out to be similar to that obtained for the extended emission, as expected if they are associated sources.

We searched for spectral variations of the photon index along the diffuse emission and found a spectral steeping with increasing from 1.6 to 2.7 with the distance to the pulsar, a behaviour observed in several Galactic PWNe, and consistent with synchrotron cooling of electrons.

\subsection{The connection with HESS J1826$-$130}
In a search for possible associations with HESS J1826$-$130, the PSR J1826$-$1256 and its nebula appear as the most plausible candidates.
In a leptonic scenario the TeV emission is expected to arise from inverse Compton (IC) scattering between the ambient low-energy photons and the same population of electron producing synchrotron radiation in the keV band. Assuming that the softening of the spectrum with the distance from the pulsar is due to cooling effects and the cosmic microwave background is the main source of background photons with a temperature of ~3 K, we roughly estimate the energy of the TeV photons produced by IC scattering. The obtained a value is lower than 30 TeV, compatible withe the detection of $\gamma$-ray in the 0.5 - 40 TeV range from HESS J1826$-$130 \citep{Anguner17}. In this way we suggest that the very high emission is likely produced by the PWN powered by PSR J1826$-$1256 via the inverse Compton mechanism.

\section{Conclusions}

The analysis of the new \textit{XMM-Newton} data confirmed the nonthermal origin for the X-ray emission with a photon index $\Gamma$ softening with the distance to the pulsar. We suggest that the most plausible origin for HESS J1826$-$130 is due to the IC mechanism within the PWN powered by PSR J1826$-$1256.

%\begin{figure}
%\center
%\includegraphics[width=\textwidth,viewport=58 38 536 331]{figure.pdf} 
%\caption{Go figure.}
%\end{figure}

% Do not delete the next line
\small  % Do not delete
%
%%% Comment the following line if you do not have acknowledgments.
\section*{Acknowledgments}   % Do not delete if you declare acknowledgments
%
%%% ACKNOWLEDGMENTS
%%% ACKNOWLEDGMENTS
L.D. is doctoral fellow of CONICET, Argentina. A.P., E.G., G.C, and L.S are members of the Carrera del Investigador Cient\'ifico of CONICET, Argentina. This work was partially supported by Argentina grants awarded by UBA (UBACyT) and ANPCYT.

%%% BIBLIOGRAPHY
\bibliographystyle{aj}
\small

\bibliography{proceedings}

\end{document}